\documentclass[journal=nalefd,manuscript=letter,layout=onecolumn]{achemso}
\setkeys{acs}{maxauthors = 0}       
\setkeys{acs}{articletitle = true} 

\usepackage[T1]{fontenc} 
\usepackage{amssymb}
\usepackage{amsmath}
\usepackage{xcolor}

\newcommand{\vect}[1]{\boldsymbol{#1}}

\author{Wei-Hua Li}
\affiliation{Department of Electrophysics, National Yang Ming Chiao Tung University, Hsinchu 300, Taiwan}

\author{Jhen-Dong Lin}
\affiliation{Department of Electrophysics, National Yang Ming Chiao Tung University, Hsinchu 300, Taiwan}

\author{Ping-Yuan Lo}
\affiliation{Department of Electrophysics, National Yang Ming Chiao Tung University, Hsinchu 300, Taiwan}

\author{Guan-Hao Peng}
\affiliation{Department of Electrophysics, National Yang Ming Chiao Tung University, Hsinchu 300, Taiwan}
									
\author{Ching-Yu Hei}
\affiliation{Department of Electrophysics, National Yang Ming Chiao Tung University, Hsinchu 300, Taiwan}

\author{Shao-Yu Chen}
\affiliation{Center of Condensed Matter Sciences, National Taiwan University, Taipei 106, Taiwan}
\affiliation{Center of Atomic Initiative for New Material, National Taiwan University, Taipei 106, Taiwan}

\author{Shun-Jen Cheng}
\affiliation{Department of Electrophysics, National Yang Ming Chiao Tung University, Hsinchu 300, Taiwan}
\email{sjcheng@nycu.edu.tw}
\title{The key role of non-local screening in the environment-insensitive exciton fine structures of transition-metal dichalcogenide monolayers}


\begin{document}
	
\begin{tocentry}
		
\includegraphics[width=\textwidth]{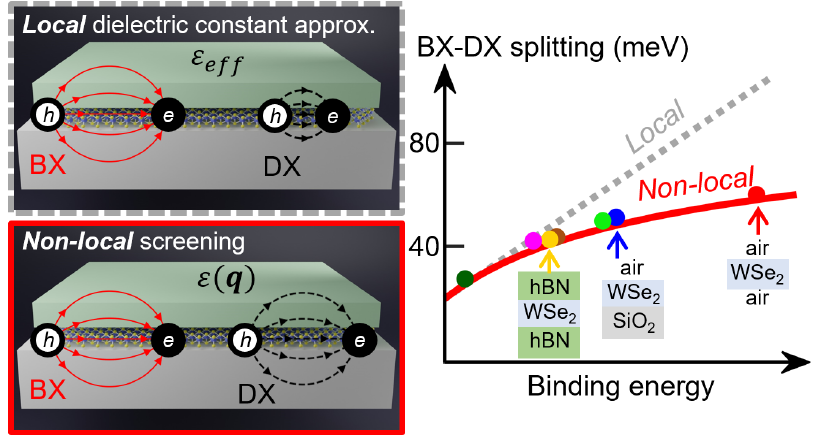}
		
\end{tocentry}
%
\begin{abstract}


In this work, we present a comprehensive theoretical and computational investigation of exciton fine structures of WSe$_2$-monolayers, one of the best known two-dimensional (2D) transition-metal dichalcogenides (TMD's), in various dielectric-layer environments by solving the first-principles-based Bethe-Salpeter equation. 
While the physical and electronic properties of atomically thin nano-materials are normally sensitive to the variation of surrounding environment, 
our studies reveal that the influence of dielectric environment on the exciton fine structures of TMD-ML’s is surprisingly limited.
We point out that the non-locality of Coulomb screening plays a key role to suppress the factor of dielectric environment and drastically shrink the fine structure splittings between bright exciton (BX) states and various dark exciton (DX) states of TMD-ML’s.
The intriguing non-locality of screening in 2D materials can be manifested by the measurable {\it non-linear} correlation between the BX-DX splittings and exciton binding energies with varying the surrounding dielectric environments. 
The revealed environment-insensitive exciton fine structures of TMD-ML's suggest the robustness of prospective dark-exciton-based opto-electronics against the inevitable variation of inhomogeneous dielectric environment.

\end{abstract}

\section{Introduction}

With the exceptional valley and excitonic properties, atomically thin transition-metal dichalcogenide monolayers (TMD-ML's) have emerged as promising nano-materials for valley-based optoelectronic and photonic applications  \cite{Coupled_Spin_Valley_PRL,xu2014spin,cao2012valley,zeng2012valley,MoS2_valley_exciton_PRB2012,
wang2015spin}.
Because of weak dielectric screening in the two-dimensional (2D) systems, photo-excited excitons in TMD-ML's are tightly bound by the enhanced electron-hole Coulomb interaction and featured with the giant binding energies ranging between 200-500 meV \cite{Nonhydrogen_Exciton_PRL2014,screening_GWBSE_PRB2016,WSe2_binding_energy,epsilon_hBN1}, leading to the superior light-matter interactions and various fascinating optical phenomena \cite{Sunlight_absorption_Nanoletters,kozawa2016evidence,wu2019ultrafast,lundt2016room}.
Even more interestingly, the pronounced Coulomb effects on those 2D excitons result in rich excitonic fine structures, which,  typically by tens of meV, well resolve the optically active bright exciton (BX) states as well as various dark exciton (DX) ones. The latter are further classified by so-called spin forbidden (SFDX) and momentum forbidden dark exciton (MFDX) states according to the violated optical selection rules \cite{OurNanoLett,Full-zone,Malic_PhysRevMaterials,deilmann2019finite,PhysRevB.105.085302,mueller2018exciton}. 
Significant fine structure splitting of exciton enables the tremendously high exciton population in the low-lying DX states and stabilizes the long-time residence of exciton \cite{SF_exp5}, giving rise to intriguing dark-exciton-related optical phenomena, such as strong optical responses in nano-photonics \cite{SF_exp6,SF_exp8,Mapara2022,su2022dark}, long-distance exciton diffusion \cite{Dark_exciton_diffusion,PhysRevLett.127.076801}, super-efficient energy transfer \cite{wu2019ultrafast}, Bose condensation \cite{DarkExciton_BEC,Combescot_2017}, etc.

In general, the physical and chemical properties of ultra-thin 2D materials such as TMD-ML's are sensitive to the variation of dielectric environments. 
Following the conventional hydrogen model of exciton in the approximation of {\it local} dielectric screening \cite{PhysRevA_2Dhydrogen,Szmytkowski2018}, the binding energy of exciton at the scale of hundreds of meV is predicted to be inversely proportional to the quadratic dielectric constant ($ \propto \varepsilon_{eff}^{-2}$), leading to the sensitive response of exciton Rydberg series to the dielectric environments  \cite{Poisson_eq,andersen2015dielectric,Diversity_trion_2017,PhysRevLett.120.057405,PhysRevB.99.115439}.
Experimentally, it has been established that the binding energy of exciton in a TMD-ML can be effectively controlled by engineering the dielectrics surrounding TMD-ML's  \cite{Raja2017,kajino2019modification,peimyoo2020engineering,stier2016probing,hsu2019dielectric} , facilitating the feasibility of exciton-based devices where charge-neutral excitons need to be transported or spatially confined.  On the other hand, it is also timely desired to find out the effective means of controlling the meV-scaled excitonic fine structures of TMD-ML's, whose BX-DX splittings directly determine the population of long lived dark exctions. 
Yet, the existing experiments reported in the literature do not observe the expected significant influences from dielectric environments on the exciton fine structures of TMD-ML's (See Table S2 of Supporting Information and references therein \cite{SF_exp1,SF_exp7,FSS_dielectric_exp,SF_exp_hBN_enc_1,Chen2018,SF_exp2,SF_exp_hBN_enc_4,SF_exp5,
SF_exp4,SF_exp9,SF_exp_hBN_enc_6,SF_exp_hBN_enc_7,SF_exp6,SF_exp3,SF_exp12,SF_exp10,SF_exp_hBN_enc_2,SF_exp_hBN_enc_8,SF_exp_hBN_enc_5,SF_exp11})  and the underlying physics remain puzzling. \cite{FSS_dielectric_exp}

In this work, we present a theoretical and computation investigation of the exciton fine structures of WSe$_2$-ML's, composed of the BX and various DX states, by solving the first-principles-based Bethe-Salpeter-equation (BSE) \cite{OurNanoLett} combined with the non-local dielectric functions of TMD-embedded multi-layer dielectric environments. 
Our studies reveal the exciton fine structure splittings of WSe$_2$-ML's under the non-local Coulomb screening are essentially weak environment-dependent, not as expected from the conventional exciton theory with local dielectric constants.
We point out the key role of non-locality of dielectric screening to suppress the dielectric-environment influences on the BX-DX fine structure splittings.
The small Bohr radii of tightly bound excitons in 2D materials suggest sizeable spreadings of the exciton wavefunctions in the momentum-space ($q$-space) and underline the essential role of $q$-dependent dielectric functions.\cite{screening_GWBSE_PRB2016,Ak_q_sapce,stier2016probing,florian2018dielectric}
Moreover, we set up an extended exciton hydrogen model with the non-local dielectric function for the TMD-ML's embedded in complex multi-layer dielectric structures. 
The theoretical studies are supported by the quantitative agreement with our experimental measurement on the exciton fine structure splittings of WSe$_2$-ML's and physically account for the commonly observed environment-insensitive dependence of exciton fine structure splittings of TMD-ML's \cite{FSS_dielectric_exp}.


\section{Results and discussion}

\subsubsection{Quasi-particle band structures of WSe$_2$ monolayer \label{sec:A}}
\sectionmark{Subsection1}

\begin{figure}[h!]
\includegraphics[width=0.85\columnwidth]{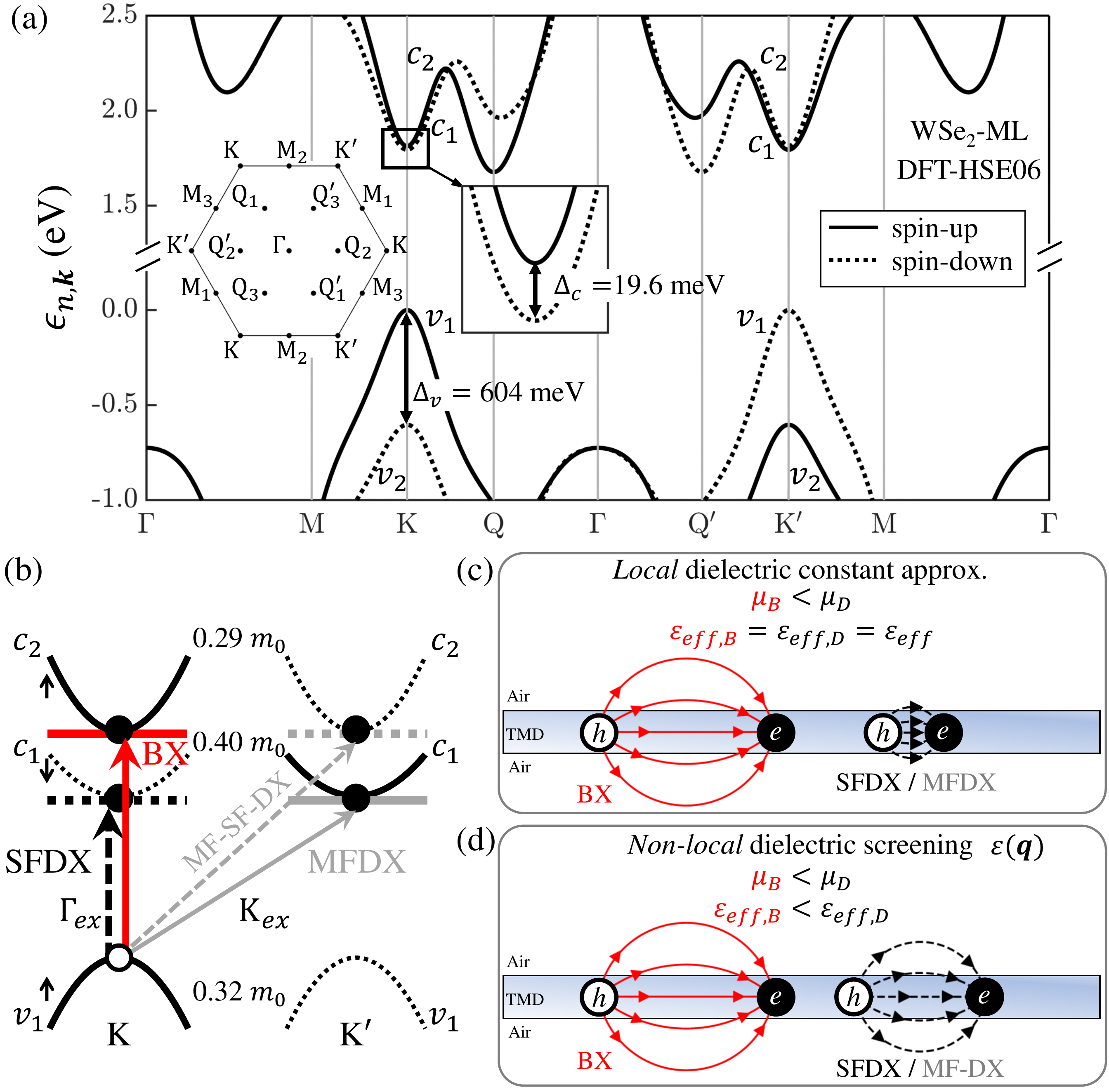}
\caption{(a) Valley- and spin-resolved quasi-particle band structure of WSe$_2$-ML calculated by using the HSE06 exchange-correlation functionals in the DFT \cite{HSE06}. 
Left inset: the first Brillouin zone of WSe$_{2}$-ML, with the indication of high symmetry points. Middle inset: the close-up of the lowest conduction bands around $K$ valley, with the spin-splitting $\Delta _{c} = 19.6$ meV. (b) Schematics of the intra-valley BX and SFDX, and the inter-valley MFDX and MF-SF-DX states, with the valence holes located at $K$ valley. 
(c) Schematics of BX and DX in a TMD-ML under a dielectric environment. In the approximation of local screening, the dielectric screening of the TMD-embedded layered system is simply characterized  by a fixed dielectric constant $\varepsilon _{eff}$, which is the same for both BX and DX. Because of heavier reduced mass of DX, the Bohr radius (binding energy) of DX is significantly smaller (larger) than the lighter BX. 
(d) Beyond the approximation of local screening, the non-local screenings for the BX and DX in the TMD-embedded layered system are described by $\vect{q}$-dependent dielectric function, $\varepsilon(\vect{q})$, leading to the unequal effective screening for the BX and DX. The non-local screening turns out to reduce the differences between the Bohr radii and binding energies of the BX and DX.} \label{Fig1}
\end{figure}

Figure~\ref{Fig1}(a) presents the quasi-particle electronic band structures of WSe$_2$ monolayer, which are calculated using the first principles Vienna Ab initio Simulation Package (VASP) \cite{VASP_package} in the density functional theory (DFT) with the use of the Heyd-Scuseria-Ernzhof (HSE) hybrid functional model \cite{HSE06} and the consideration of spin-orbit coupling (SOC) (See Supporting Information for the technical details). As this work is focused on the low-lying spectra of A-exciton, we shall pay the main attention on the low-lying conduction and topmost valence bands, as shown by Fig.~\ref{Fig1}(a). Due to the effect of SOC, the conduction [valence] bands at the $K$ and $K'$ valleys are split by $\Delta_c (K/K^{\prime}) \equiv \epsilon_{c_2, K/K^{\prime}} - \epsilon_{c_1, K/K^{\prime}}  \approx 19.6 \text{ meV} >0 $ [$\Delta_v (K/K^{\prime}) \equiv  \epsilon_{v_1, K/K^{\prime}} - \epsilon_{v_2, K/K^{\prime}}  \approx 604$ meV], which leads also to the distinct effective masses of the spin-split conduction bands in the $K$ and $K'$ valleys as indicated in Fig.~\ref{Fig1}(b) and Table \ref{Table1}. Since the lowest spin-split conduction band possesses the opposite spin to that of the topmost valence band in the same valley, the lowest exciton states of WSe$_2$-ML are expected to be SFDX states. 
In addition to $\Delta_c$, the Coulomb interactions, both direct and exchange ones, of exciton substantially affect the BX-DX fine structure splittings of WSe$_2$-ML's, as will be shown later.

\subsection{Theory of exciton fine structures of TMD-ML's under dielectric screenings \label{sec:B}}
\sectionmark{Subsection2}

To compute the exciton spectra, we employ the theoretical methodology developed by Ref.[\citenum{OurNanoLett}] that establishes the Bethe-Salpeter equation (BSE) on the first-principles base and numerically solves the exciton fine structures in an efficient manner. The BSE governing the exciton wave function and energy spectra comprises the kinetic energies of electron-hole pairs and the {\it e-h} Coulomb kernel that consists of the attractive {screened} {\it e-h} direct interaction and the repulsive exchange one.\cite{OurNanoLett} The screened direct Coulomb interaction is associated with the non-local dielectric function of a TMD-ML surrounded by specific dielectric environment that can be solved by using classical electro-magnetic theory. 
The precise evaluation of the matrix elements of the Coulomb kernel that essentially involves the rapidly varying microscopic parts of the Bloch wave functions is computationally non-trivial.  We take the strategy of wannierizing the Bloch wave functions into the maximally localized Wannier functions by using the package Wannier90  \cite{MOSTOFI2008685,MOSTOFI20142309} and employ the Wannier basis to reformulate the Kohn-Sham equation in the matrix form of manageable dimensions. Taking the approach, the evaluation of the electron-hole exchange interactions in terms of Wannier functions can be carried out with significantly reduced numerical cost.  \cite{OurNanoLett}

Following Ref.[\citenum{Poisson_eq}], the 2D Fourier transform of the screened direct Coulomb interaction $W \left( \boldsymbol{q} ; z _{1}, z _{2} \right)$ is obtained by solving the Poisson's equation, which reads
\begin{equation}\label{Poisson}
\left[- q^{2} \varepsilon_{\boldsymbol{q}} (z_{1}) + \frac{\partial}{\partial z_{1}} \varepsilon _{\boldsymbol{q}} (z_{1}) \frac{\partial}{\partial z_{1}} \right] W(\boldsymbol{q};z_{1},z_{2}) = -\frac{e^{2}}{\varepsilon _{0}} \delta(z_{1}-z_{2}) \, ,
\end{equation}
where $\vect{q}=(q_x,q_y)$ is the in-plane wave vector, $z$ is the out-of-plane coordinate,  $\varepsilon_{\boldsymbol{q}} \left( z \right)$ is the $z$-dependent macroscopic dielectric function for the dielectric system consisting of a TMD-ML and the surrounding dielectric layers.  For the dielectric layers, $\varepsilon_{\boldsymbol{q}} \left( z \right)$ is considered to be piecewise constants (See Eq.(S6) of Supporting Information).  For the sandwiched TMD-ML itself, we adopt the non-local dielectric functions and the parameters for WSe$_2$-ML given by Refs.[\citenum{Bulk_TMD_dielectric, PhysRevLett.124.257402}].

\begin{figure}[h!]
\includegraphics[width=0.9\columnwidth]{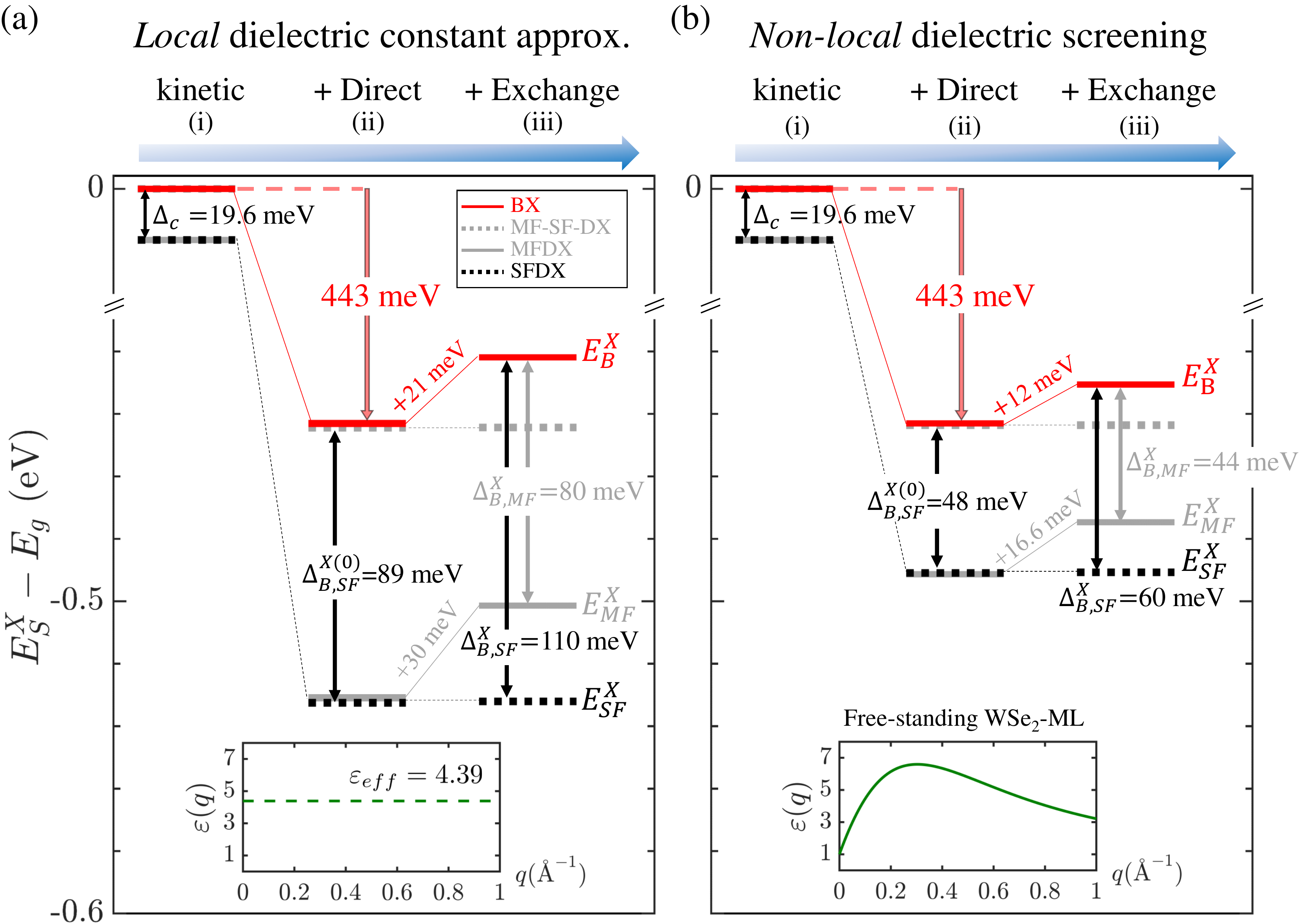}
\caption{The BSE-calculated low-lying exciton fine structures of free-standing WSe$_{2}$-ML (a) with fixed $\varepsilon_{eff} = 4.39$ in the approximation of local screening and (b) with the $\vect{q}$-dependent non-local dielectric function, $\varepsilon(\vect{q})$ (See the inset at the bottom), beyond the approximation of local screening. (i)-(iii) in (a) and (b): The exciton fine structure spectra calculated by solving the modified or exact BSE's that take into accounts (i) only the kinetic energies, (ii) the kinetic energies and the direct Coulomb interaction and (iii) all the kinetic energies and both the direct and exchange Coulomb interactions, respectively.  From the evolution of the spectra from (i), (ii) to 
(iii), the {\it e-h} direct interaction is realized to be the dictating mechanism of the BX-SFDX splitting. Comparing (a) and (b), the non-local screening is shown to make the overall shrinkage of the fine structure splittings between the BX and various DX's. } \label{Fig2}
\end{figure}

By solving Eq.~(\ref{Poisson}), we can solve the effective non-local dielectric functions for various TMD-ML-embedded multi-layer systems with the number of dielectric layers up to five (See Eq. (S15)). Taking the $z$-average of $W \left( \boldsymbol{q} ; z _{1} , z _{2} \right)$ over the thickness of TMD-ML \cite{Poisson_eq}, the Fourier transform of the $z$-average of the screened Coulomb interaction is given by $W \left( \boldsymbol{q} \right) \approx \frac{1}{d ^{2}} \int _{- d / 2} ^{d / 2} \int _{- d / 2} ^{d / 2} d z _{1} d z _{2} \, W \left( \boldsymbol{q} ; z _{1}, z _{2} \right)$.
Likewise,
$V \left( \vect{q} \right) = \, \frac{1}{d^{2}}  \int _{- d / 2} ^{d / 2} d z _{1} \int _{- d / 2} ^{d / 2} d z _{2} \int d ^{2} \vect{\rho} \, V \left( \vect{\rho} ; z _{1} - z _{2} \right) e ^{- i \vect{q} \cdot  \vect{\rho} }  =  \frac{e ^{2}}{4 \pi \varepsilon _{0} } \, \frac{4 \pi}{d q ^{2}} \left( 1 - \frac{1 - e ^{- q d}}{q d} \right)$
is the $z$-average of bare Coulomb interaction. Accordingly, one can evaluate the non-local dielectric function defined by $\varepsilon \left( \boldsymbol{q} \right) =  V \left( \boldsymbol{q} \right) / W \left( \boldsymbol{q} \right)  $ \cite{epsilon_hBN1,epsilon_q_VW}. 
Incorporating the solved $\varepsilon \left( \boldsymbol{q} \right)$ for a free-standing WSe$_2$-ML, {\it{i.e.}}, the case (I) of Fig.\ref{Fig3}, in the Coulomb kernel of BSE, the exciton fine structure of a free-standing WSe$_2$-ML under non-local screening is numerically calculated by solving the BSE and shown in Fig.~\ref{Fig2}(b).

To confirm the validity of the theoretical prediction, we experimentally perform the cryogenic photoluminescence (PL) measurement on the sample of hBN-encapsulated WSe$_{2}$-ML (corresponding to the case V of Fig.\ref{Fig3}) and compare the measured optical spectrum with the BSE-calculated fine structure of BX and DX's. The experimental details are given in Supporting Information. The computed exciton fine structures with the full consideration of the Coulomb interactions and the non-local dielectric function show the splitting between the BX and SFDX states of hBN-encapsulated WSe$_{2}$-ML to be 42.7 meV, in excellent agreement with the measured BX-SFDX splitting, 41.8 meV, shown in Fig.S2. In addition to the case of hBN-encapsulated WSe$_{2}$-ML, the exciton fine structures of WSe$_{2}$-ML's with different surrounding dielectric materials are also calculated and shown consistent with the measured splittings ever reported in the literature as summarized by Table S2.

\subsection{Exciton fine structures in the approximation of {\it local} screening \label{sec:C}}
\sectionmark{Subsection3}

For a systematic investigation, we first purposely disregard the non-locality of screening and replace the {\it non-local} $q$-dependent dielectric function with a fitted effective dielectric constant, $\varepsilon_{eff}$, in the Coulomb kernel of BSE.
In the approximation of local screening, Fig.~\ref{Fig2}(a) shows the calculated low-lying fine structure of exciton for a free-standing WSe$_2$-ML with the effective dielectric constant $\varepsilon_{eff}=4.39$ in the modified BSE that takes into account (i) only the kinetic energies,  (ii) the kinetic energies and screened direct Coulomb interaction, and (iii) the kinetic energies and the full Coulomb interactions including the both direct and exchange ones, respectively. 
The effective dielectric constant $\varepsilon_{eff}= 4.39$ is determined by fitting the numerically calculated binding energy of BX, $E ^{b}_{B}= 443$ meV, for a free-standing WSe$_2$-ML to the $\varepsilon_{eff}$-parametrized 2D hydrogen model.\cite{PhysRevA_2Dhydrogen,Szmytkowski2018}

In the 2D hydrogen model of exciton, the energy of exciton in the $S$-state ($S=B, SF, MF, ...$) is explicitly given by  $E_{S}^{X(0)} = E_g - 4Ry(\frac{\mu_S}{m_0}) \varepsilon_{eff}^{-2} $, with the binding energy of exciton,
\begin{equation} \label{Eb}
E_S^b \equiv E_{g} - E^{X(0)}_{S} = 4Ry(\frac{\mu_S}{m_0}) \varepsilon_{eff}^{-2} \, , 
\end{equation}
where $Ry=13.6$ eV is the Rydberg constant and $\mu_S=(m_c^{-1} + m_v^{-1} )^{-1}$ is the reduced mass of exciton. 
For WSe$_2$-ML, $\mu_{B}=0.15 \, m_{0}$ and $\mu_{SF}=0.18\, m_{0}$ (See Table \ref{Table1}).

Taking only the kinetic energies of electron and hole, the non-interacting exciton fine structures shown in Fig.~\ref{Fig2}(a-i) are simply featured with two multiple-fold degenerate levels split by  $\Delta_c = 19.6 $ meV. 
Turning on the direct Coulomb interaction as shown in Fig.~\ref{Fig2}(a-ii), the degeneracies of the exciton states remain the same but the BX-SFDX splitting $\Delta_{B, SF}^{X(0)}\equiv E_{B}^{X(0)} - E_{SF}^{X(0)}$ is significantly enlarged, from 19.6 meV increased to 89 meV.
 Hereafter, we use the superscript $(0)$ to denote the exchange-free exciton states subjected only to the attractive {\it e-h} direct interaction and the abbreviated subscripts $B$, $SF$ and $MF$ to denote BX, SFDX, and MFDX, respectively. 
The unequal reduced masses of BX and SFDX, $\mu_{B} < \mu_{SF} $, result in the different Bohr radii, $ a_{B}^{X} > a_{SF}^{X}$, which are defined by $a_S^{X} \equiv \left( \frac{4\pi\epsilon_0 \hbar^2}{e^2} \right) \cdot \left( \frac{\varepsilon_{eff} }{\mu_S } \right) $.
For a free-standing WSe$_2$-ML with $\varepsilon_{eff}=4.39$, $a_{B}^{X}=1.5$ nm ($a_{SF}^{X}=1.3$ nm) for BX (SFDX) with the reduced mass, $\mu _B=0.15 \;  m_0$ ($\mu_{SF}=0.18 \;m_0$) is determined.
With the smaller Bohr radius, the heavier SFDX gains more {\it e-h} Coulomb attraction and acquires larger exciton binding energy than the lighter BX (See Fig.\ref{Fig1}(c) for schematic illustration). 
According to Eq.~(\ref{Eb}), the unequal binding energies of the BX and SFDX states ($E_B^b < E_{SF}^b$) increase the BX-SFDX energy splitting by 
\begin{equation} \label{FSS_direct}
\Delta_{B,SF}^{X(0)} = \Delta _{c} + \frac{4Ry}{m_0 \varepsilon_{eff}^2}\cdot  \left( \mu_{SF} - \mu_{B} \right)\  ,
\end{equation}
evaluated as 89 meV, which is several times of magnitude greater than $\Delta_c=19.6$ meV.

Further taking the {\it e-h} exchange interaction into account, as shown in Fig.~\ref{Fig2} (a-iii), the energy of BX states is raised by the short-ranged (SR) {\it e-h} exchange interaction, numerically calculated as $ V_{B}^{x} = E_{B}^{X}-E_{B}^{X(0)} = 21$ meV, while the energy of the lowest SFDX states free of exchange interaction remains unchanged. 
Thus, under the approximation of {\it local} screening with the fixed $\varepsilon_{eff}$, the BX-SFDX splitting is predicted to be $ \Delta_{B, SF}^{X} \equiv E_{B}^{X} - E_{SF}^{X}= 110$ meV, which is however much larger than the measured splitting in the range of $\sim 50$ meV  \cite{SF_exp1,SF_exp2,SF_exp3,SF_exp4,SF_exp5,SF_exp6,SF_exp7,SF_exp8,SF_exp9,SF_exp10,SF_exp11,SF_exp12,SF_exp13}. 
In fact, as will be shown later, the non-locality in dielectric screening neglected so far is essential in the exciton fine structures, which drastically reduces the splitting to fall into the experimental range.

\begin{table}[t]
\begin{tabular}{|cccc|cccccc|}
\hline
\multicolumn{4}{|c|}{Effective masses of carriers ($m_{0}$)} & \multicolumn{6}{c|}{ Reduced masses of excitons ($m_{0}$)}  \\
\hline
$m_{v_{1},K}$ & $m_{c_{1},K}$ & $m_{c_{2},K}$ & $m_{c_{1},Q}$ & {} & {} & $\mu _{B}$ & {} & {} & $\mu _{SF}$  \\
 0.32 & 0.40 & 0.29 & 0.51 & {} & {} & 0.15 & {} & {} & 0.18  \\
\hline
\end{tabular}
\caption{Effective masses of the conduction electrons and valence holes in the $K$- or $Q$-valleys, the reduced masses of bright exciton ($\mu_B$) and that of spin-forbidden dark exciton ($\mu_{SF}$) in the unit of free electron mass $m_{0}$.}
\label{Table1}
\end{table}

\subsection{Fine structure of exciton under {\it non-local} dielectric screening \label{sec:D}}
\sectionmark{Subsection4}

Figure~\ref{Fig2}(b) shows the calculated exciton fine structure of a free standing WSe$_{2}$-ML by solving the BSE with the {\it non-local} dielectric function, $\varepsilon(\vect{q})$, shown by the lower inset in Fig.~\ref{Fig2}(b). In comparison with Fig.~\ref{Fig2}(a-iii), the BX-DX splitting, $\Delta_{B,SF}^{X}$, of the free standing WSe$_{2}$-ML under the {\it non-local} dielectric screening shown in Fig.~\ref{Fig2}(b-iii) is drastically decreased from 110 meV to 60 meV. Overall, the fine structure splittings are shrunk by the decreases of the both direct- and exchange-interaction-induced splittings due to the non-local effect of screening.
The decrease of the BX-DX splitting results because the Bohr radius (binding energy) of the heavier DX states is increased (decreased) by the non-local screening more than the lighter BX states, as illustrated by Fig.\ref{Fig1}(c) and (d).  In other words, the effective dielectric constant of the DX states turns out to be greater than that of BX, {\it{i.e.}}, $\varepsilon_{eff,D}= \varepsilon_{eff,B} + \Delta \varepsilon $ with  $\Delta \varepsilon >0 $, as considering the non-locality of screening.

With a given $\boldsymbol{q}$-dependent dielectric function, $\varepsilon (\boldsymbol{q})$, the effective dielectric constant for an exciton in the $S$ state  ($S=B, SF ,MF, ...$) can be estimated by \cite{epsilon_eff_Thygesen}
\begin{equation}\label{epsilon_eff_Thygesen}
\varepsilon_{eff,S} \approx 2 \left( \frac{a^{X}_{S}}{2} \right) ^{2} \int _{0}^{2/a^{X}_{S}} dq \; q  \varepsilon (q) \, .
\end{equation}
 The formalism of Eq.~(\ref{epsilon_eff_Thygesen}) suggested by Ref.[ \citenum{epsilon_eff_Thygesen}] is based on the fact that the major part of the exciton wave function is localized in the finite reciprocal  $\vect{q}$-space roughly enclosed by the circle of the radius $\sim 2/a^{X}_{S}$. The large binding energy of an exciton in TMD-ML's leads to a small Bohr radius of a few nm, suggesting a significant spreading of the exciton wavefunction in the $\vect{q}$-space. Therefore, a $\vect{q}$-dependent non-local dielectric function becomes critical to capture the microscopic difference between different kinds of excitonic complexes. 

To determine $\Delta \varepsilon$ in an analytical manner, we extend the conventional hydrogen model combined with the non-local dielectric function, 
\begin{equation} \label{epsilon_series}
\varepsilon(q) \approx c_{0} + c_{1} q + c_{2} q^2 + ... \, ,
\end{equation}
expanded in a Taylor series for small $q \in \{ 0, 2/a_S^X\}$, where $c_{0}\equiv \varepsilon(q)\vert_{q=0}$, $c_{1} \equiv \partial \varepsilon(q)/\partial q \vert_{q=0}$ and $c_{2} \equiv \frac{1}{2} \partial^2 \varepsilon(q)/\partial q^2 \vert_{q=0}$.   
Taking $a^{X}_{S}=\left( \frac{4\pi\varepsilon_0 \hbar^2}{e^2}\right) \left( \frac{ \varepsilon_{eff,S}}{\mu_S} \right) $  as a function of $\varepsilon_{eff,S}$ in Eq.~(\ref{epsilon_eff_Thygesen}), one can solve Eq.~(\ref{epsilon_eff_Thygesen}) for $\varepsilon_{eff,S}$ in a iterative manner.

In the first-order approximation, where $\varepsilon(q) \approx c_{0} + c_{1} q $,  the difference between  $\varepsilon_{eff,D}$ and $\varepsilon_{eff,B}$ is derived from Eq.~(\ref{epsilon_eff_Thygesen}) as
\begin{align}\label{Delta_eps}
\Delta \varepsilon \equiv  \varepsilon_{eff,D}- \varepsilon_{eff,B}  & = \frac{1}{2} \sqrt{ c_{0}^2 + \frac{4e^2 c_1}{3\pi \varepsilon_0 \hbar^2}\mu_{D}  } -  \frac{1}{2} \sqrt{ c_{0}^2 + \frac{4e^2 c_1}{3\pi \varepsilon_0 \hbar^2}\mu_{B}  }\, . 
\end{align}
With $\mu_{D} >  \mu_{B}$, Eq.~(\ref{Delta_eps}) shows $\Delta \varepsilon  > 0 $, accounting for the greater effective dielectric constant of the heavier DX than the lighter BX.

For better quantitative studies, one can expand $\varepsilon(q) \approx c_{0} + c_{1} q +c_{2} q^2$ up to the second-order term, which is shown sufficient to yield the results in good agreement with the numerical ones as shown in Fig.\ref{Fig4}. For symmetric three-layer structures, {\it{i.e.}}, a TMD-ML sandwiched by two semi-infinite dielectrics, the $q$-dependent dielectric functions expanded in Taylor series are solvable and considered in the model analysis throughout this work (See Supporting Information for more details).  In the the second-order approximation, $\Delta \varepsilon = 0.18 $ is estimated for a free-standing WSe$_2$-ML. Taking $\Delta \varepsilon = 0.18 $, the calculated  BX-SFDX splitting of a free-standing WSe$_{2}$-ML is decreased from  122 meV to 62 meV, in agreement with the numerically calculated splitting that is 60 meV.

\subsection{Dielectric-environment-dependences of exciton fine structures \label{sec:E}}
\sectionmark{Subsection5}

\begin{figure}[p!]
\includegraphics[width=0.7\columnwidth]{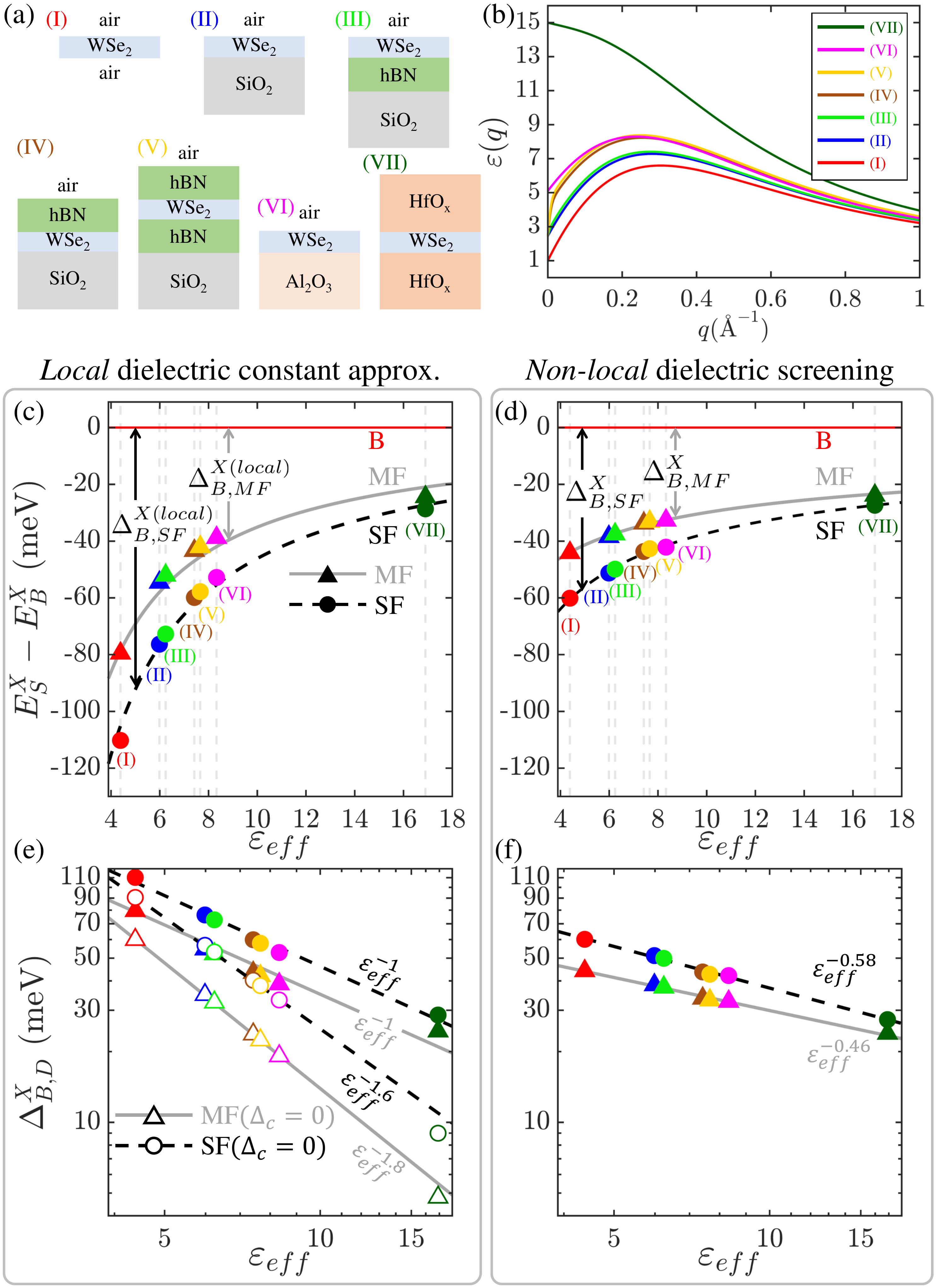}
\caption{(a) The WSe$_{2}$-ML-embedded dielectric environments considered in this work.  In the cases of (I)-(VII), all the substrate and capping layers are considered semi-infinitely thick, except for the hBN layer whose thickness is set to be 10 nm.  (b) The calculated $\vect{q}$-dependent dielectric functions for the dielectric systems (I)-(VII). (c) The calculated relative energy levels of the BX, MFDX and SFDX states of the WSe$_{2}$-ML under the dielectric screenings of (I)-(VII), indicated by filled triangles (MFDX) and circles (SFDX) in different colors, in the approximation of local screening with the fitted dielectric constants, $\varepsilon_{eff}$. Dark dashed and gray solid curves are for eye-guiding. (d) is the same as (c) but under the non-local dielectric screening described by $\varepsilon(\vect{q})$.  (e) [(f)]: The log-log plot for the BX-DX splittings, $\Delta_{B,D}^X = E_{B}^X - E_{D}^X$, given by (c) [(d)], fitted by the power-law of $\Delta_{B,D}^X \propto \varepsilon_{eff}^{-n}$ with the exponent $n$. The white-filled symbols in (e) are the calculated splittings by disregarding $\Delta _{c}$. Taking into account the non-local screening, the exponent is decreased to the small values of $n\sim 0.58, 0.46$, indicating the weak environment-dependence of the BX-DX splitting.
} \label{Fig3}
\end{figure}

Now, let us investigate the influence of dielectric environment on the exciton fine structures of TMD-ML's with the full consideration of non-local dielectric screening. 
Table S2 of Supporting Information summarizes from the literature the experimentally measured fine structure splittings between BX and SFDX states of WSe$_{2}$ monolayers in the dielectric environments of different materials and structures, which consistently fall in a narrow energy range between 40-50meV for the various dielectric environments and show the weak environment-dependences.\cite{FSS_dielectric_exp}
For realistic simulations, we consider WSe$_2$-ML's under various complex dielectric environments in the multi-layer structures composed of  (I)  air/WSe$_{2}$/air (free-standing TMD-ML), (II) air/WSe$_{2}$/SiO$_{2}$, (III) air/WSe$_{2}$/hBN/SiO$_{2}$, (IV) air/hBN/WSe$_{2}$/SiO$_{2}$, (V) air/hBN/WSe$_{2}$/hBN/SiO$_{2}$, (VI) air/WSe$_{2}$/Al$_{2}$O$_{3}$ and (VII) HfO$_{x}$/WSe$_{2}$/HfO$_{x}$, as depicted in Fig.~\ref{Fig3}(a). In the layered structures of (I)-(VII), all the substrate and capping layers are considered semi-infinitely thick, except for the hBN layer whose thickness is set to be 10 nm. The dielectric constants of the dielectric materials of the layer structures (I)-(VII) are shown in Table.S1.

Figure~\ref{Fig3}(d) shows the exciton fine structure spectra of BX, inter-valley MFDX and intra-valley SFDX states of the WSe$_2$-ML's in the dielectric environments of Fig.~\ref{Fig3}(a) calculated by solving the BSE with the full non-local dielectric functions $\varepsilon(\vect{q})$, versus the fitted effective dielectric constants $\varepsilon _{eff}$ for the screened TMD-systems. 
By fitting the numerically calculated binding energy of BX to the formalism of $\varepsilon _{eff}$-parametrized hydrogen model, $\varepsilon _{eff}$ = 4.39, 5.99, 6.24, 7.42, 7.66, 8.33 and 16.9, are determined for the cases (I)-(VII) of Fig.~\ref{Fig3}(a), respectively.
To underline the non-local effect of screening, Fig.~\ref{Fig3}(c) shows the calculated exciton fine structure spectra of the dielectric-surrounded WSe$_2$-ML's by solving the BSE with fixed constant of $\varepsilon _{eff}$ for comparison. 
  
Figure~\ref{Fig3}(e) is the log-log plot, based on the data of Fig.~\ref{Fig3}(c) in the approximation of local screening, of the calculated BX-DX splittings of the screened WSe$_2$-ML's versus $\varepsilon _{eff}$. It is shown that the calculated BX-DX splittings of WSe$_2$-ML under local screening follow the $\varepsilon_{eff}$-power law, $\Delta_{B,D=SF,MF}^X \propto \varepsilon_{eff}^{-n}$, with the exponent $n \approx 1$.
The value of the exponent $n\sim 1$ is smaller than that of the power-law for the exciton binding energy, {\it{i.e.}}, $n = 2$ as shown by Eq.~(\ref{Eb}), because the $\varepsilon_{eff}$-independent $\Delta_c$ takes part in the BX-DX splitting. Artificially removing $\Delta_c$ from the calculation, the $\Delta_c$-free BX-SFDX (BX-MFDX) splitting shows the $\varepsilon_{eff}$-power law with the $n\approx 1.6$ ($n\approx 1.8$), very close to $n=2$ given by the conventional hydrogen model.

Figure~\ref{Fig3}(f) is the log-log plot of the calculated BX-DX splittings of WSe$_2$-ML's based on the data of Fig.~\ref{Fig3}(d) with the consideration of non-local screening, showing the power-law of $\Delta_{B,SF}^X $ in $ \varepsilon_{eff}$ with $n \approx 0.58$. The very small value of $n$ indicates the relatively weak $\varepsilon_{eff}$-dependences of the BX-DX splitting of WSe$_2$-ML's and account for the experimental observations of Table S2 and Ref.[\citenum{FSS_dielectric_exp}].
The weak $ \varepsilon_{eff}$-dependence of BX-DX splitting is explicitly shown by Eq.(S28) as a consequence of non-local dielectric screening, which decreases the BX-DX splitting by the cubic $ \varepsilon_{eff}^{-3}$-term, $\Delta_{B,D}^{X(non-loc.)} = -8 \varepsilon_{eff}^{-3} \Delta \varepsilon (\mu_D/m_0) Ry$, associated with the non-zero $\Delta \varepsilon$ that reflects $\mu_D \neq \mu_B$ as illustrated by Fig.\ref{Fig1}(c).

\begin{figure}[p!]
\includegraphics[width=0.9\columnwidth]{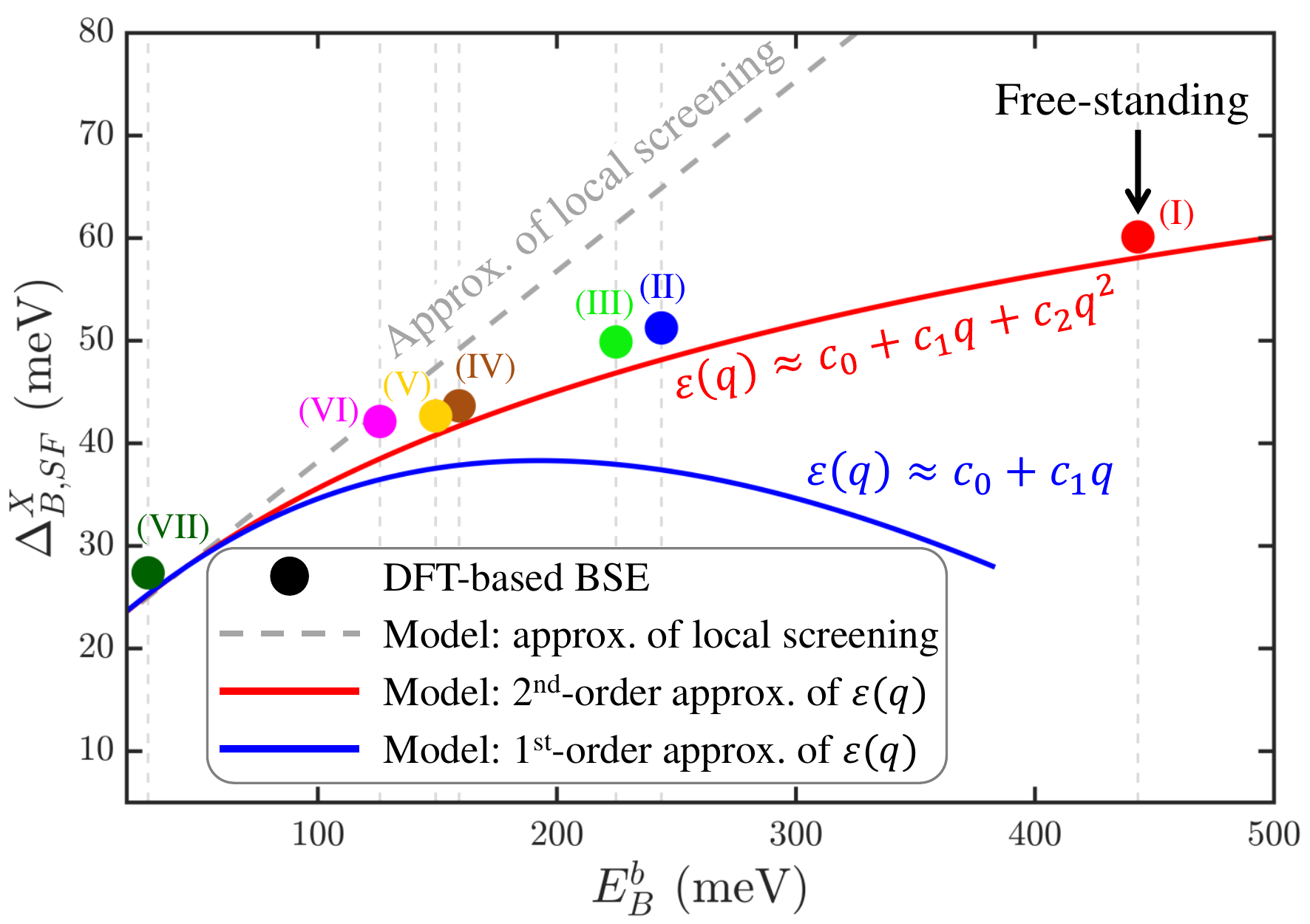}
\caption{The non-linear correlation between the BX-SFDX splittings ($\Delta_{B,SF}^X$) and binding energies of BX ($E_B^b$) of WSe$_{2}$-ML with changing the dielectric environments from (I) to (VII) of Fig.~\ref{Fig3}(a). The coloured circles are obtained by solving DFT-based BSE for the WSe$_{2}$-ML's in the multi-layer dielectric structures with the full consideration of the non-local dielectric functions. The gray dashed line is the predicted linear $\Delta_{B,SF}^X$-$E_B^b$ by the conventional hydrogen model in the approximation of local screening. The red and blue solid lines are the non-linear correlation simulated by the extended hydrogen model with the expanded non-local dielectric functions in the 1st- and 2nd-order approximations of $\varepsilon (q)$, respectively. 
The sub-linear correlation between $\Delta_{B,SF}^X$ and $E_B^b$ manifests the insensitive environment-dependence of exciton fine structures as the consequence of non-local dielectric screening.}
\label{Fig4}
\end{figure}

\subsection{Non-linear correlation between the BX-DX splitting and exciton binding energy under non-local screening \label{sec:F}}
\sectionmark{Subsection6}

The non-locality in dielectric screening can be manifested by the non-linear correlation between the BX-DX splitting and binding energy with varying $\varepsilon_{eff}$.
In the approximation of local screening, Eqs.~(\ref{Eb}) and (\ref{FSS_direct}) show that the BX-DX splitting and the binding energy of an exciton in a TMD-ML is dominated by the $\varepsilon_{eff}^{-2}$ terms. 
Beyond the approximation of local screening, the non-locality-induced $\varepsilon_{eff}^{-3}$ term with negative sign in Eq.(S28) takes part in the BX-DX splitting and makes non-linear the correlation between the BX-DX splitting and exciton binding energy.

Figure~\ref{Fig4} presents the scatter plot of the numerically BSE-calculated BX-SFDX splittings versus the exciton binding energies of WSe$_{2}$-ML's in the multi-layer dielectric structures (I)-(VII) shown in Fig.~\ref{Fig3}(a) with the consideration of the non-local dielectric functions. As expected from the model analysis,  the BSE-calculated fine structure splittings and binding energies of exciton with varying $\varepsilon_{eff}$ are shown non-linearly correlated and follow a sub-linear relationship.

Beyond the approximation of local screening, the extended hydrogen model combined with the approximate non-local $\varepsilon(q)$ expanded up to the first-order term in Eq.~(\ref{epsilon_series}) predicts the blue curve in Fig.~\ref{Fig4}. It is clearly seen that, by introducing the $q$-dependence of dielectric function, the  BX-SFDX splitting becomes weakly correlated with the binding energy of exciton. Yet, compared with the numerical data, the model simulation in the first-order approximation for $\varepsilon(q)$ overestimates the non-linear correlation between the BX-SFDX splitting and binding energy of exciton.
The validity of the exciton model with non-local screening can be improved by taking the non-local dielectric function expanded up to the second-order term and solve the $\Delta \varepsilon$ more precisely [See Eq.~(S21)]. 
The model-simulated result with the non-local dielectric function in the second-order approximation shows an excellent agreement with the numerically BSE-calculated data.

\section{Conclusion}
In summary, we carry out a comprehensive theoretical investigation, implemented on the first-principles base, of the exciton fine structures of WSe$_{2}$-ML under the dielectric screening from complex dielectric-layer environments.
While the physical and electronic properties of atomically thin 2D materials are generally sensitive to the variation of surrounding environment, our first-principles-based theoretical studies justified by the agreement with the measured optical spectra point out that the non-locality of dielectric screening plays a key role to suppress the influence of dielectric environment on the exciton fine structure spectra of 2D materials. The theoretical studies account for the weak environment-dependences of exciton fine structure splittings of TMD-ML's commonly observed in the existing experiments. In addition to the fully numerical studies, we set up an extended exciton hydrogen model with the non-local dielectric function successfully enabling the quantitative simulation and analysis for the exciton fine structures of TMD-ML's surrounded by complex multi-layer dielectrics.
The intriguing non-locality in dielectric screening is shown to be manifested by the measurable non-linear correlation between the BX-DX splittings and exciton binding energies with varying the effective dielectric constant of environment.
Despite the poor tunability of exciton fine structure by engineering dielectric environment, the revealed weak environment-dependences of exciton fine structures suggest the robustness of dark-exciton-based physical properties against the variation or fluctuation of dielectric environment.

\section{Methods}

\subsection{Sample fabrication and experimental photoluminescence (PL) measurement}
\sectionmark{Subsection_methods_1}

We exfoliated WSe$_{2}$ from the bulk WSe$_{2}$ crystal purchased from HQ Graphene. The WSe$_{2}$-ML flakes are firstly exfoliated on a PDMS stamp and inspected under optical microscope. The thickness of WSe$_{2}$ is confirmed by optical contrast and optical spectroscopy. Then, we transferred the flakes on the desired substrates ({\it{e.g.}}, SiO$_{2}$, Al$_{2}$O$_{3}$, {\it{etc.}}) For hBN-encapsulated WSe$_{2}$-ML samples, we followed the fabrication procedures of PPC-assisted transfer mentioned in Ref.[\citenum{Chen2018}]. The top and bottom hBN are chosen around 20 nm to ensure good passivation and protection.

After fabrication, the samples were mounted in a Janis (ST-500) cryostat maintained at a high vacuum ($5 \times 10^{-6}$ torr) and cooled down to 10 K by flowing liquid helium. For PL spectroscopy, we excited the sample by a CW laser with wavelength at 532 nm. We employed a $40 \times$ objective lens (NA: 0.6) to focus the laser to a spot size of about $\sim 1$ $\mathrm{\mu m}$ on the sample and to collect the PL signal in back-scattering geometry. The signal was guided to the Horiba iHR 550 spectrometer and then detected by the liquid-nitrogen cooled charge-coupled device of Horiba Symphony II detection system.

\subsection{First-principles calculations and Bethe-Salpeter equation (BSE) for exciton spectra}
\sectionmark{Subsection_methods_2}

The quasi-particle states of WSe$_{2}$-ML in DFT are carried out by utilizing the Vienna Ab initio Simulation Package (VASP) \cite{VASP_package} with the Heyd-Scuseria-Ernzhof (HSE) exchange-correlation functional \cite{HSE06}. We employ the experimental lattice constant 3.282 $\text{\AA}$ \cite{ALHILLI197293,SCHUTTE1987207} in the DFT calculation. The energy cutoff of the plane-wave basis is set to be 500 eV for the expansion of the wavefunctions and the projector-augmented wave (PAW) potentials. The gamma-centered k-grid of $9 \times 9 \times 1$ is taken to sample the Brillouin zone and the electronic relaxation convergence limit is set as $10^{-6}$ eV. In order to simulate the suspended WSe$_{2}$ monolayer, the distance between the periodic layers in the out-of-plane direction is set to be greater than 30 $\text{\AA}$. The structure relaxation is performed with the convergence limit of 0.005 $\text{eV/\AA}$.

The exciton spectra are carried out by the theoretical methodology developed by Ref.[\citenum{OurNanoLett}] that establishes Bethe-Salpeter equation (BSE) based on the DFT-calculated quasi-particle band structures of 2D materials. Following the methodology used in Ref.[\citenum{OurNanoLett}], we utilize the package Wannier90 \cite{MOSTOFI2008685,MOSTOFI20142309} to transform the DFT-calculated Bloch functions into the set of maximally localized Wannier functions (MLWFs) and reconstruct the Kohn-Sham Hamiltonian in the wannier tight-binding scheme. Thus, the matrix elements of the Coulomb kernel of BSE are evaluated in terms of MLWFs obtained by utilizing the package Wannier90 \cite{MOSTOFI2008685,MOSTOFI20142309}. Incorporating with the interpolation techniques \cite{OurNanoLett}, the evaluation of the matrix elements of the Coulomb kernel can be carried out with high accuracy and low numerical cost.


\begin{acknowledgement}
	

This study is supported by the National Science and Technology Council of Taiwan, under contracts, MOST 109-2112-M-009-018-MY3, and by National Center for High-Performance Computing (NCHC), Taiwan.

\end{acknowledgement}

\begin{suppinfo}
	

	
The Supporting Information is available free of charge at http://pubs.acs.org/doi/xxxxxxx

\begin{itemize}
  \item  Theory for the non-local dielectric function of layered structure, summary of the dielectric constants of capping and substrate materials (Table.S1), experimental PL spectrum of hBN-encapsulated WSe$_{2}$-ML (Fig.S2), techniques of the sample fabrication and PL measurement, technical details of DFT calculations, summary of the measured BX-DX splittings of WSe$_2$ monolayers in various dielectric environments reported in the literature (Table.S2) and the extended hydrogen model of exciton with non-local dielectric screening (PDF)
\end{itemize}

\end{suppinfo}


\bibliography{screening_refs}

\end{document}